\documentclass[prb,twocolumn,showkeys,showpacs,amsmath,amsfonts,floatfix,superscriptaddress,nofootinbib]{revtex4-1}

\usepackage{amssymb,amsmath,amstext}                
\usepackage{graphicx}                                               
\usepackage{epstopdf}                                               
\usepackage{color}       
\usepackage{bm}
\usepackage{appendix}                                              
\usepackage[latin2]{inputenc}
\usepackage{ulem}
\usepackage{latexsym}
\usepackage[colorlinks=true,citecolor=blue,linkcolor=magenta]{hyperref}
\newcommand{\bra}[1]{\mbox{$\langle #1 |$}}
\newcommand{\ket}[1]{\mbox{$| #1 \rangle$}}
\newcommand{\braket}[2]{\mbox{$\langle #1  | #2 \rangle$}}

\def\be{\begin{equation}}
\def\ee{\end{equation}}
\def\bea{\begin{eqnarray}}
\def\eea{\end{eqnarray}}
\def\bi{\begin{itemize}}
\def\ei{\end{itemize}}

\begin{document}

\title{ Time evolution of an infinite projected entangled pair state: \\
             a gradient tensor update in the tangent space }

\author{Jacek Dziarmaga}
\email{dziarmaga@th.if.uj.edu.pl}
\affiliation{Jagiellonian University, Institute of Theoretical Physics, 
             ul. \L{}ojasiewicza 11, 30-348 Krak\'ow, Poland }

\date{May 23, 2022}

\begin{abstract}
Time evolution of an infinite 2D many body quantum lattice system can be described by the Suzuki-Trotter decomposition applied to the infinite projected entangled pair state (iPEPS). Each Trotter gate increases the bond dimension of the tensor network, $D$, that has to be truncated back in a way that minimizes a suitable error measure. This paper goes beyond simplified error measures -- like the one used in the full update (FU), the simple update (SU), and their intermediate neighborhood tensor update (NTU) -- and directly maximizes an overlap between the exact iPEPS with the increased bond dimension and the new iPEPS with the truncated one. The optimization is performed in a tangent space of the iPEPS variational manifold. This gradient tensor update (GTU) is benchmarked by a simulation of a sudden quench of a transverse field in the 2D quantum Ising model and the quantum Kibble-Zurek mechanism in the same 2D system. 
\end{abstract}

\maketitle

\section{ Introduction }
\label{sec:introduction}

In typical condensed matter applications quantum many body states can be represented efficiently by tensor networks~\cite{Verstraete_review_08,Orus_review_14}. They include the one-dimensional (1D) matrix product state (MPS)~\cite{fannes1992}, its two-dimensional (2D) version called a projected entangled pair state (PEPS)~\cite{Nishino_2DvarTN_04,verstraete2004}, or a multi-scale entanglement renormalization ansatz~\cite{Vidal_MERA_07,Vidal_MERA_08,Evenbly_branchMERA_14,Evenbly_branchMERAarea_14}. MPS is a compact representation of ground states of 1D gapped local Hamiltonians \cite{Verstraete_review_08,Hastings_GSarealaw_07,Schuch_MPSapprox_08} and purifications of their thermal states \cite{Barthel_1DTMPSapprox_17}. It is the ansatz optimized by the density matrix renormalization group (DMRG) \cite{White_DMRG_92, White_DMRG_93,Schollwock_review_05,Schollwock_review_11}. By analogy, though with some reservations \cite{Eisert_TNapprox_16}, PEPS is expected to be suitable for ground states of 2D gapped local Hamiltonians \cite{Verstraete_review_08,Orus_review_14} and their thermal states \cite{Wolf_Tarealaw_08,Molnar_TPEPSapprox_15,Alhambra2021}. As a variational ansatz tensor networks do not suffer from the sign problem common in the quantum Monte Carlo and they can deal with fermionic systems \cite{Corboz_fMERA_10,Eisert_fMERA_09,Corboz_fMERA_09,Barthel_fTN_09,Gu_fTN_10,Cirac_fPEPS_10,Corboz_fiPEPS_10,Corboz_stripes_11}.

\begin{figure}[t!]
\vspace{-0cm}
\includegraphics[width=0.99999\columnwidth,clip=true]{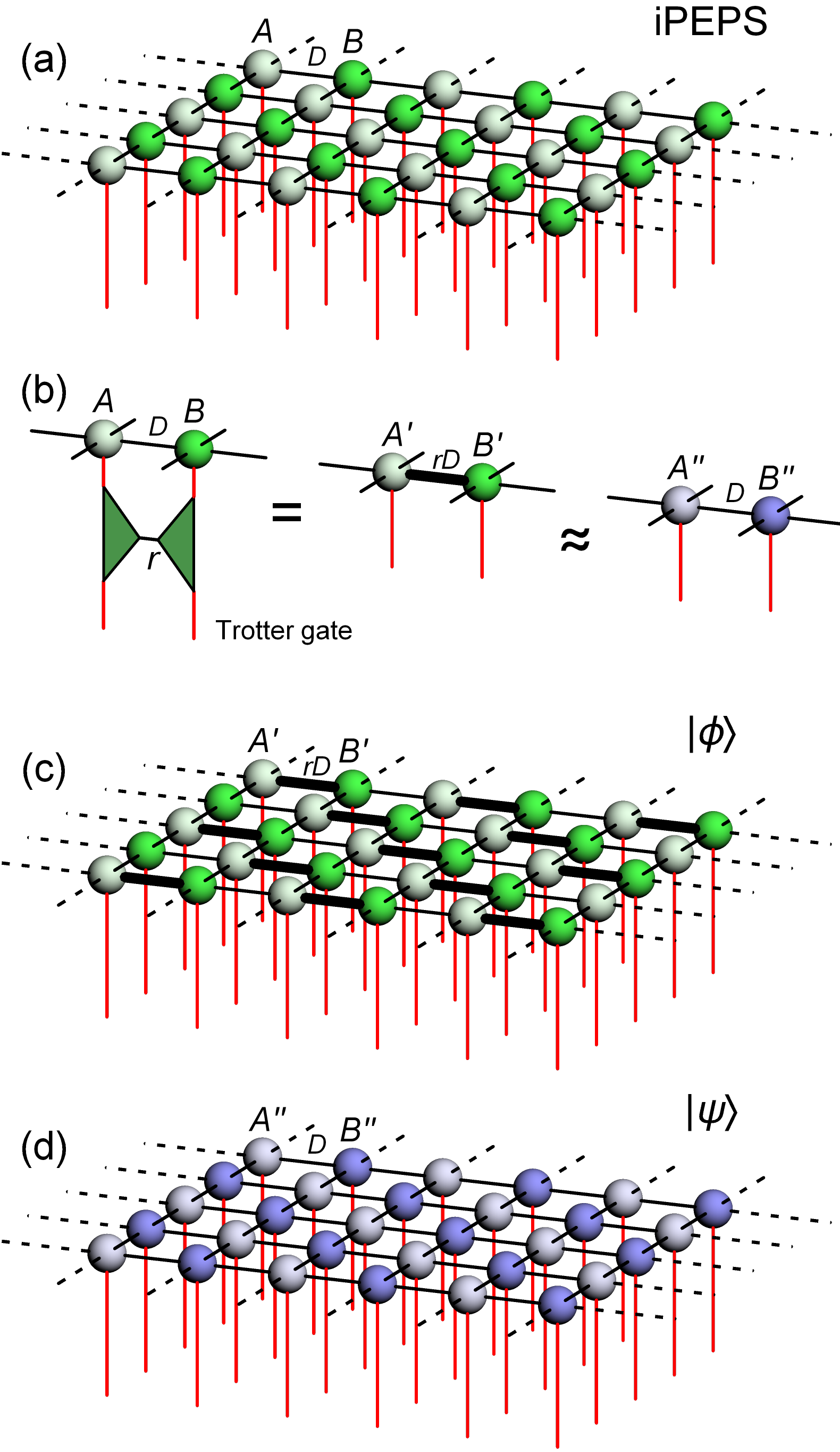}
\vspace{-0cm}
\caption{
{\bf Trotter gate. }
An iPEPS is an inifinite checkerboard of tensors $A$ and $B$ in (a). Every tensor has one physical index (red) and four bond indices of dimension $D$ (black) connecting it with its nearest neighbors.
In (b) the iPEPS is applied a Trotter gate at every horizontal bond between $A$ and $B$.
The gate is a made of two tensors contracted by an index of dimension $r$.
Its application increases the size of the bond index to $r D$.
The initial iPEPS (a), with the Trotter gate applied at every horizontal bond $A-B$, 
becomes an exact state $\ket{\phi}$ in (c). 
In order to reduce the bond dimension back to $D$ 
the middle diagram in (b) has to be approximated by the right diagram with a pair of new tensors: $A^{''}-B^{''}$. 
The new tensors make new iPEPS $\ket{\psi}$ in (d).
$A^{''}$ and $B^{''}$ are optimized to maximize an overlap between the exact state and the new iPEPS:  $\vert\braket{\phi}{\psi}\vert^2/\braket{\psi}{\psi}$.
}
\label{fig:2site}
\end{figure}

Originally proposed as an ansatz for ground states of finite systems~\cite{Verstraete_PEPS_04, Murg_finitePEPS_07,Nishino_2DvarTN_04}, subsequent development of efficient numerical methods for infinite systems \cite{Cirac_iPEPS_08,Xiang_SU_08,Gu_TERG_08,Orus_CTM_09} promoted infinite PEPS (iPEPS), shown in Fig. \ref{fig:2site}(a), to one of the methods of choice for strongly correlated systems in 2D. It was crucial for solving the long-standing magnetization plateaus problem in the highly frustrated compound $\textrm{SrCu}_2(\textrm{BO}_3)_2$ \cite{matsuda13,corboz14_shastry}, establishing the striped nature of the ground state of the doped 2D Hubbard model \cite{Simons_Hubb_17}, and new evidence supporting gapless spin liquid in the kagome Heisenberg antiferromagnet \cite{Xinag_kagome_17} (though tensor renormalization group suggests gapped spin liquid with a long correlation length \cite{TRG_kagome_Wen}). Recent developments in iPEPS optimization \cite{fu,Corboz_varopt_16,Vanderstraeten_varopt_16}, contraction \cite{Fishman_FPCTM_17,Xie_PEPScontr_17}, energy extrapolations~\cite{Corboz_Eextrap_16}, and universality-class estimation \cite{Corboz_FCLS_18,Rader_FCLS_18,Rams_xiD_18} opened an avenue towards even more challenging problems, including simulation of thermal states \cite{Czarnik_evproj_12,Czarnik_fevproj_14,Czarnik_SCevproj_15,Czarnik_compass_16,Czarnik_VTNR_15,Czarnik_fVTNR_16,Czarnik_eg_17,Dai_fidelity_17,CzarnikDziarmagaCorboz,czarnik19b,Orus_SUfiniteT_18,CzarnikKH,wietek19,jimenez20,CzarnikSS,Poilblanc_thermal}, mixed states of open systems~\cite{Kshetrimayum_diss_17,CzarnikDziarmagaCorboz,SzymanskaPEPS}, excited states \cite{Vanderstraeten_tangentPEPS_15,ExcitationCorboz}, or unitary evolution \cite{CzarnikDziarmagaCorboz,HubigCirac,tJholeHubig,Abendschein08,SUlocalization,SUtimecrystal,ntu,KZ2D,BH2Dcorrelationspreading,mbl_ntu}. 

The unitary evolution is the subject of this paper. As in previous work we adapt the Suzuki-Trotter decomposition \cite{Trotter_59,Suzuki_66,Suzuki_76} of the evolution operator into a product of nearest neighbor(NN) Trotter gates. As before each Trotter gate increases the bond dimension that has to be truncated in order to prevent its exponential growth. However, we do not rely on local optimization like the simple update (SU) \cite{tJholeHubig,SUlocalization}, the full update (FU) \cite{fu,CzarnikDziarmagaCorboz}, or the neighbourhood tensor update (NTU) \cite{ntu,KZ2D,mbl_ntu}, but employ further gradient optimization to directly maximize an overlap between the truncated iPEPS and the exact one with the increased bond dimension. The optimization operates in the tangent space of the iPEPS variational manifold. The Gramm-Schmidt metric tensor and the gradient are obtained with the corner transfer matrix renormalization group \cite{corboz14_tJ,corboz16b}.

This paper is organized as follows. In Sec. \ref{sec:gradient} we introduce the gradient optimization after the exact iPEPS is pre-truncated with NTU. In Sec. \ref{sec:gradient} we outline calculation of the metric tensor and the gradient in the tangent space of the iPEPS. More details on the use of reduced tensors/matrices instead of full iPEPS tensors can be found in appendix \ref{app:reduced}. In Sec. \ref{sec:sudden} the gradient tensor update (GTU) method is subject to the standard benchmark of a sudden quench in the 2D transverse field quantum Ising model and in Sec. \ref{sec:kz} by the Kibble-Zurek ramp in the same system. Truncation errors during GTU evolution are presented in appendix \ref{app:errors}. We conclude in Sec. \ref{sec:conclusion}.

\section{ Tangent space optimization }
\label{sec:gradient}

The iPEPS is an inifinite checkerboard of tensors $A$ and $B$ in Fig. \ref{fig:2site} (a).
In a Suzuki-Trotter step a two-site Trotter gate is applied to every equivalent NN bond between sites/tensors $A$ and $B$. The bond dimension on the applied bonds is increased by a factor of $r$ where $r$ is a SVD rank of the gate. We refer to this new iPEPS as $\ket{\phi}$. Our goal is to approximate this exact state with an iPEPS, $\ket{\psi}$, where all bond dimensions are brought back to the original $D$. The latter iPEPS is made of tensors $A^{''}$ and $B^{''}$ which are to be treated as variational parameters. Tensors $A^{''}$ and $B^{''}$ are optimized in a loop, $\to A^{''}\to B^{''}\to$, until convergence of the overlap between $\ket{\phi}$ and $\ket{\psi}$. In the following we explain one step of the loop when $A^{''}$ is optimized at fixed $B^{''}$. The other step, when $B^{''}$ is optimized at fixed $A^{''}$, is done in a similar way.

We consider small variations of the state, $\ket{\psi}\to\ket{\psi}+\ket{\delta\psi}$, due to variations of the tensor: $A^{''}\to A^{''}+\delta A^{''}$, which are orthogonal to $\ket{\psi}$. Up to linear order in $\delta A^{''}$ the variation of the state is
\be 
\ket{\delta\psi}=
\left(
1-\frac{\ket{\psi}\bra{\psi}}{\langle \psi | \psi \rangle}
\right)
\sum_\mu \delta A^{''}_\mu ~ \ket{\partial^\mu\psi}.
\label{deltapsi}
\ee 
Here index $\mu$ numbers elements of tensor $\delta A^{''}$ and $\partial^\mu$ is a derivative with respect to element $A^{''}_\mu$. $\delta A^{''}$ has to minimize a cost function:
\bea 
F
&=&
\left|
\ket{\phi}-\ket{\psi}-\ket{\delta\psi}
\right|^2 /
\langle \psi | \psi \rangle
\nonumber\\
&=&
\delta A^{''*}_\mu ~ G^{\mu\nu} ~ \delta A^{''}_\nu - \delta A^{''*}_\mu ~ J^\mu - J^{\mu *} ~ \delta A^{''}_\mu + F_0.
\label{F}
\eea 
Here the repeated indices imply summation, the Gramm-Schmidt metric is 
\bea
G^{\mu\nu} 
&=&
\langle \psi | \psi \rangle^{-1}
\bra{\partial^\mu\psi}
\left(
1-\frac{\ket{\psi}\bra{\psi}}{\langle \psi | \psi \rangle}
\right)
\ket{\partial^\nu\psi}
\nonumber\\
&=&
\frac{\langle \partial_\mu \psi | \partial_\nu \psi \rangle}{\langle \psi | \psi \rangle}-
\frac{\langle \partial_\mu \psi | \psi \rangle}{\langle \psi | \psi \rangle}
\frac{\langle \psi | \partial_\nu \psi \rangle}{\langle \psi | \psi \rangle}
\label{G}
\eea 
and the gradient
\bea
J^\mu
&=&
\langle \psi | \psi \rangle^{-1}
\bra{\partial^\mu\psi}
\left(
1-\frac{\ket{\psi}\bra{\psi}}{\langle \psi | \psi \rangle}
\right)
\left(
\ket{\phi}-\ket{\psi}
\right)
\nonumber\\
&=&
\frac{\langle \psi | \phi \rangle}{\langle \psi | \psi \rangle}
\left[
\frac{\langle \partial^\mu \psi | \phi \rangle}{\langle \psi | \phi \rangle}-
\frac{\langle \partial^\mu \psi | \psi \rangle}{\langle \psi | \psi \rangle}
\right]
\nonumber\\
&\approx &
\frac{\langle \partial^\mu \psi | \phi \rangle}{\langle \psi | \phi \rangle}-
\frac{\langle \partial^\mu \psi | \psi \rangle}{\langle \psi | \psi \rangle}
\label{J}
\eea 
In the last approximate equality we assume $\langle \psi | \phi \rangle \approx \langle \psi | \psi \rangle$ that becomes accurate close to convergence. The quadratic cost function \eqref{F} is minimized by
\be 
\delta A^{''}_\mu = G_{\mu\nu} J^\nu. 
\label{steepest}
\ee 
Here $G_{\mu\nu}$ is a (pseudo-)inverse of $G^{\mu\nu}$. 

In order to go beyond the linear approximation in $\delta A^{''}_\mu$ in \eqref{deltapsi} we use the solution \eqref{steepest} to construct a new variational iPEPS $\ket{\psi_x}$ where all $A^{''}$ are replaced by
\be 
A^{''} + x ~ \left[ \delta A^{''} - A^{''} \left( J^{\mu *} \delta A^{''}_\mu \right) \right]
\label{Ax}
\ee 
and $x$ is a real variational parameter. Note that to linear order in small $x$ it satisfies:
$
\ket{\psi_x} \approx \ket{\psi}+x \ket{\delta\psi},
$ 
hence $x$ parametrizes the line of steepest descend of the cost function \eqref{F} which is tangent to the iPEPS manifold at $\ket{\psi}$. Following this line promises the fastest reduction of $F$. However, beyond the quadratic approximation in the second line of \eqref{F}, valid for small $x$, it is better to use a logarithmic overlap:
\be 
O_x = 
\left(     
\frac{ \langle \psi_x | \phi \rangle \langle \phi | \psi_x \rangle }
     { \langle \psi_x | \psi_x \rangle }
\right)^{1/N},
\ee 
where $N\to\infty$ is a number of lattice sites. The overlap does not suffer form the orthogonality catastrophe for large $N$ and is computable by tensor network methods. In the following calculations the linear search algorithm was used to optimize the overlap with respect to $x$. With the optimal $x$ we accept \eqref{Ax} as new $A^{''}$ and proceed to optimize $B^{''}$ in a similar manner. The optimization of tensors $A^{''}$ and $B^{''}$ is repeated in a loop until convergence.

\begin{figure}[t!]
\vspace{-0cm}
\includegraphics[width=0.9999\columnwidth,clip=true]{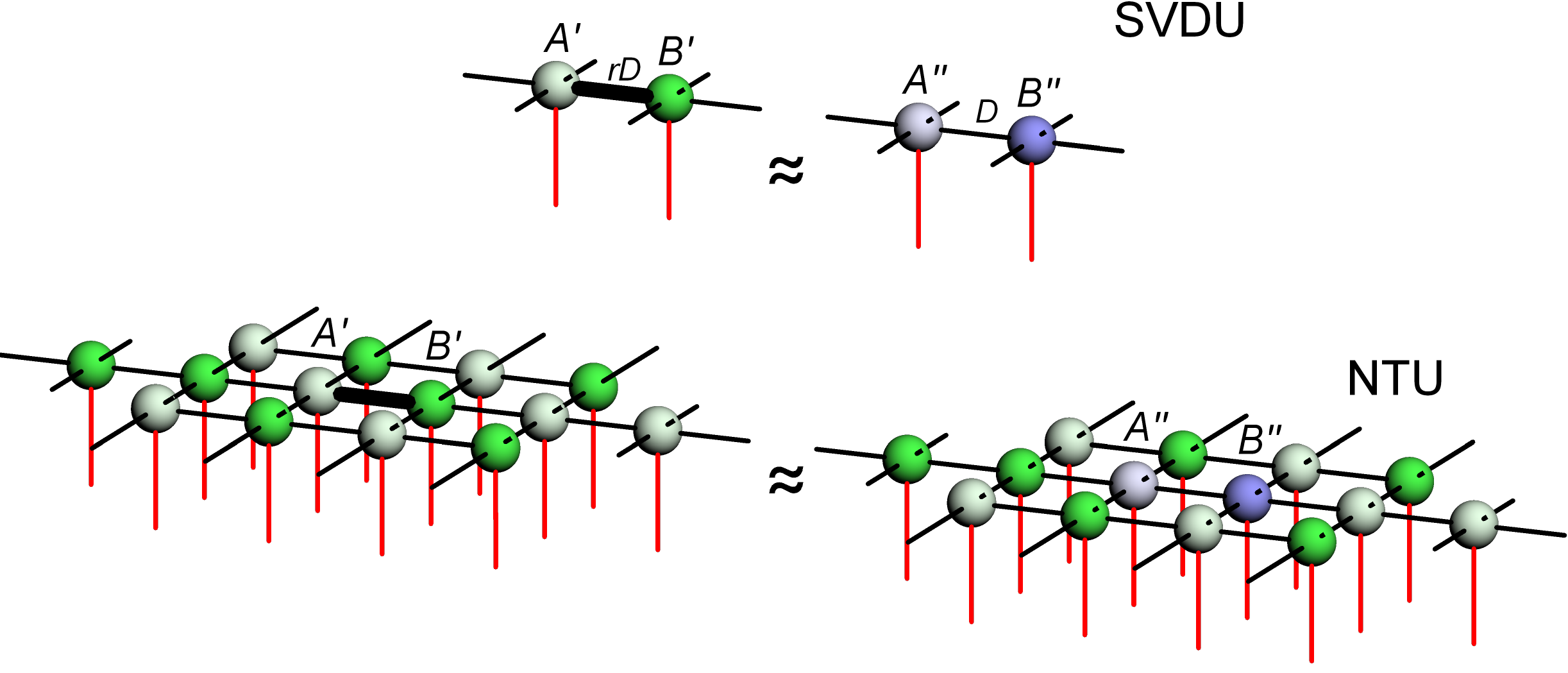}
\vspace{-0cm}
\caption{
{\bf Initialization. }
Before the gradient optimization loop new tensors $A^{''}$ and $B^{''}$ are initialized with the neighbourhood tensor update (NTU) \cite{ntu}. 
In NTU $A^{''}$ and $B^{''}$ are optimized variationally to minimize a Frobenius norm of a difference between the left and the right diagrams in the bottom panel. 
NTU in turn is initialized with $A^{''}$ and $B^{''}$ obtained by SVD truncation of the thick $rD$-bond in the top panel.  
This SVD update (SVDU) minimizes the difference between the two diagrams in the top panel. 
Note that the NTU clusters in the bottom are larger than in Ref. \onlinecite{ntu}.
}
\label{fig:ntu}
\end{figure}

In order to avoid trapping by local minima, initial $A^{''}$ and $B^{''}$ for the gradient optimization are obtained by a robust neighbourhood tensor update \cite{ntu} (NTU) outlined in Fig. \ref{fig:ntu}. NTU in turn is initialized by a simple SVD truncation of the $rD$-bond that we call SVD update (SVDU). Thus in fact the optimization of $A^{''}$ and $B^{''}$ after each Trotter gate proceeds in three stages:
\be 
{\rm SVDU} \longrightarrow {\rm NTU} \longrightarrow {\rm GTU}.
\ee 
At the end of each stage quality of the approximation is monitored by an overlap per site:
\be 
O = 
\left(     
\frac{ \langle \psi | \phi \rangle \langle \phi | \psi \rangle }
     { \langle \psi | \psi \rangle }
\right)^{1/N},
\label{O}
\ee 
where $\ket{\psi}$ is the best iPEPS obtained after the SVDU/NTU/GTU stage.

\section{ The metric and the gradient }
\label{sec:gradient}

\begin{figure}[t!]
\vspace{-0cm}
\includegraphics[width=1.0\columnwidth,clip=true]{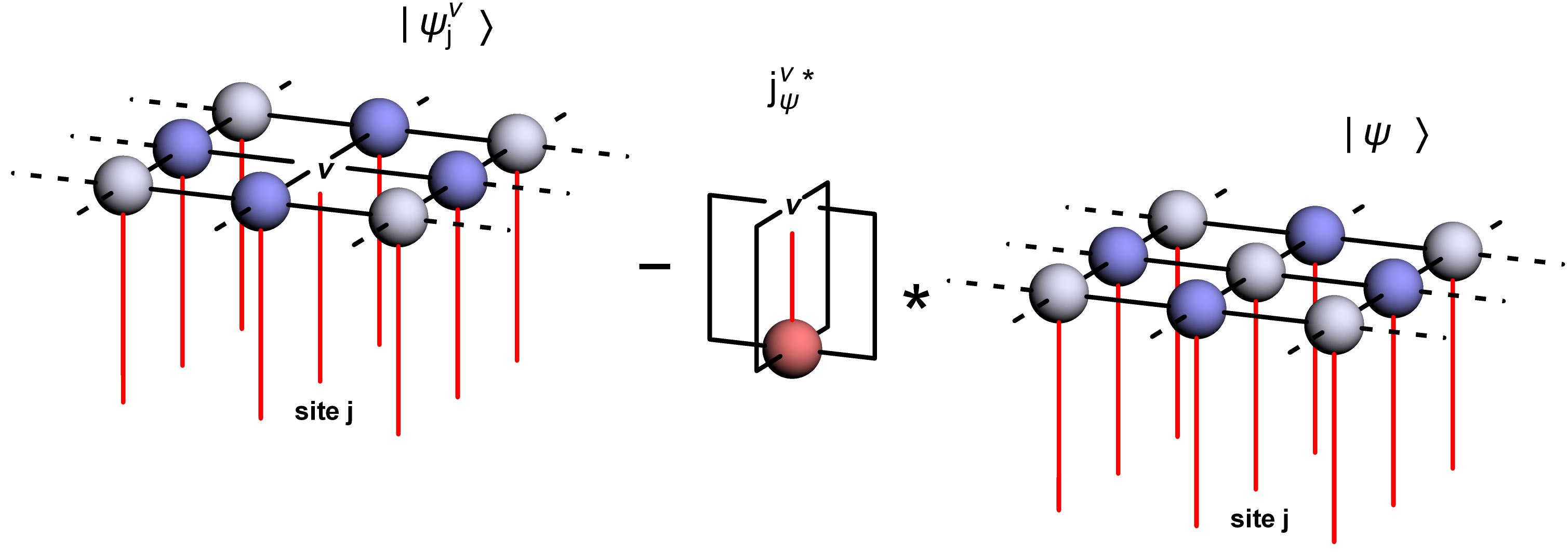}
\vspace{-0cm}
\caption{
{\bf Connected iPEPS derivative.} 
$| \psi_j^\nu \rangle_{\rm c}$ in \eqref{G0} is equal to $| \psi_j^\nu \rangle - j^{\nu *}_\psi | \psi \rangle$.
A derivative of new iPEPS $\ket{\psi}$ with respect to tensor element $A^{''}_\nu$ is a sum over index $j$ -- numbering sites on sublattice A -- of tensor networks $\ket{\psi_j^\nu}$. 
Here $\ket{\psi_j^\nu}$ is iPEPS $\ket{\psi}$ with one tensor $A^{''}$ missing at site $j$. The removal of $A^{''}$ creates open indices -- four bond and one physical -- that take the set of values $\nu$.
The subtraction makes it orthogonal to the new iPEPS: 
$\braket{\psi}{\psi_j^\nu}_{\rm c}=0$. 
}
\label{fig:ren}
\end{figure}

In this section we outline calculation of the Gramm-Schmidt metric tensor \eqref{G} and the gradient \eqref{J} with the corner transfer matrix renormalization group \cite{corboz14_tJ,corboz16b}. To begin with, notice that 
\be 
\ket{\partial^\nu\psi} = \sum_s \ket{\psi_s^\nu},
\label{derivative}
\ee 
where $\ket{\psi_s^\nu}$ is the same as iPEPS $\ket{\psi}$ except that one tensor $A^{''}$ located at site $s$ is missing, see the leftmost diagram in Fig. \ref{fig:ren}. The free indices take the set of values $\nu$. Here index $s$ runs over sublattice $A$ only. Accordingly, the gradient in \eqref{J} becomes
\bea  
J^\mu & = & 
\sum_s
\frac{\langle \psi_s^\mu | \phi \rangle}{\langle \psi | \phi \rangle}-
\frac{\langle \psi_s^\mu | \psi \rangle}{\langle \psi | \psi \rangle}
\nonumber \\
&=&
N 
\left(
\frac{\langle \psi_0^\mu | \phi \rangle}{\langle \psi | \phi \rangle}-
\frac{\langle \psi_0^\mu | \psi \rangle}{\langle \psi | \psi \rangle}
\right)
\nonumber\\
&\equiv&
N 
\left(
j^\mu_\phi - j^\mu_\psi
\right)
\label{J0}
\eea  
Here $0$ is a label for a single reference site in the infinite lattice. Evaluation of \eqref{J0} can be done with the corner transfer matrix renormalization group (CTMRG) in the same way as for a 1-site expectation value \cite{corboz14_tJ}.

In a similar manner the Gramm-Schmidt metric in \eqref{G} becomes
\bea
G^{\mu\nu} 
&=&
\sum_{s,s'}
\frac{ \langle \psi_s^\mu | \psi_{s'}^\nu \rangle }{ \langle \psi | \psi \rangle }-
\frac{ \langle \psi_s^\mu | \psi \rangle }{ \langle \psi | \psi \rangle }
\frac{ \langle \psi | \psi_{s'}^\nu \rangle }{ \langle \psi | \psi \rangle }
\nonumber \\
&=&
N
\frac{ \langle \psi_0^\mu | }{ \langle \psi | \psi \rangle }
\sum_s
| \psi_s^\nu \rangle -
| \psi \rangle
\frac{ \langle \psi | \psi_s^\nu \rangle }{ \langle \psi | \psi \rangle }
\nonumber \\
&=&
N
\frac{ \langle \psi_0^\mu | }{ \langle \psi | \psi \rangle }
\sum_s
| \psi_s^\nu \rangle -
j^{\nu *}_\psi
| \psi \rangle
\nonumber\\
&=&
N 
\frac{ \langle \psi_0^\mu | }{ \langle \psi | \psi \rangle }
\sum_s
| \psi_s^\nu \rangle_{\rm c}
\label{G0}
\eea 
Here $| \psi_s^\nu \rangle_{\rm c} \equiv | \psi_s^\nu \rangle - j^{\nu *}_\psi | \psi \rangle$ is a connected derivative of new iPEPS with respect to $A^{''}_\nu$ at site $s$. The substraction on the RHS makes it orthogonal to the new iPEPS: 
$\braket{\psi}{\psi_j^\nu}_{\rm c}=0$. Thanks to the orthogonality the sum in \eqref{G0} has non-zero contributions only from sites $j$ that are within a correlation range from the reference site $0$. 
The sum can be done with CTMRG in the same way as for a connected correlation function \cite{corboz16b}.

Last but not least, in order to make the calculations computationally efficient, in place of full tensors $A^{''}$ and $B^{''}$ we optimize their reduced tensors/matrices, as explained in appendix \ref{app:reduced}. 

\begin{figure}[t!]
\vspace{-0cm}
\includegraphics[width=1.0\columnwidth,clip=true]{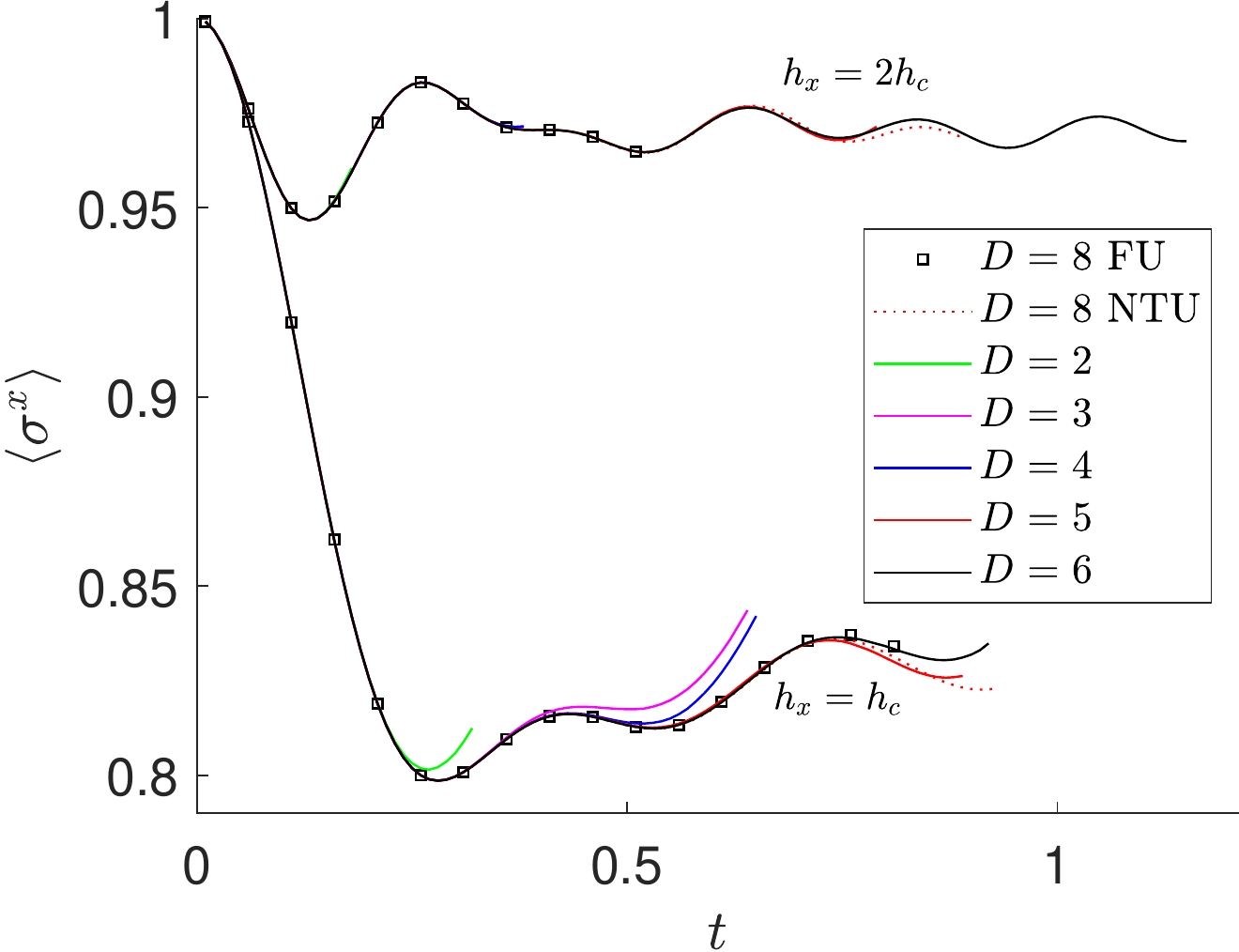}
\caption{
{\bf Transverse magnetization after a quench.} 
The unitary evolution of $\langle \sigma^x \rangle$ after a sudden quench from a fully polarized state. Two bunches of curves are shown corresponding to evolution with $h_x=2h_c,h_c$. Each curve is terminated when the energy per site departed by $0.01$ from its initial value or the error measured by overlap falls below a threshold. The squares are data from FU simulations \cite{CzarnikDziarmagaCorboz} and extend as long as they appear converged for $D=8$. The dashed red line comes from NTU simulations \cite{ntu} and are terminated when the energy departs by $0.01$. Here we use the same time step, $dt=0.01$, and the same second order Suzuki-Trotter decomposition as in FU and NTU. 
}
\label{fig:x}
\end{figure}

\section{Evolution after a sudden quench}
\label{sec:sudden}

Here as in Refs. \cite{CzarnikDziarmagaCorboz,ntu} we consider a sudden quench in the transverse field quantum Ising model on an infinite square lattice:
\be 
H =
-J\sum_{\langle j,j'\rangle} \sigma^z_j \sigma^z_{j'} - h_x \sum_j \sigma^x_j.
\label{HI}
\ee
At zero temperature the model has a ferromagnetic phase with non-zero spontaneous magnetization $\langle \sigma^z \rangle$ for the magnitude of the transverse field, $|h_x|$, below a quantum critical point located at $h_c=3.04438(2)$\cite{Deng_QIshc_02}. 

We set $J=1$ and simulate unitary evolution after a sudden quench at time $t=0$ from infinite transverse field down to a finite $h_x$. After $t=0$ the fully polarized ground state of the initial Hamiltonian is evolved by the final Hamiltonian with $h_x=2h_c,h_c$. The same quenches were simulated with tensor networks \cite{CzarnikDziarmagaCorboz,ntu} in the thermodynamic limit and neural quantum states \cite{ANN_Markus&Markus} on a finite lattice. They are probably the most challenging application for a tensor network simulation because the sudden quench of the Hamiltonian creates lots of excitations, especially the quench to the gapless quantum critical point. As far as one can think in terms of quasiparticles, they are created as entangled pairs with opposite quasimomenta. By moving in opposite directions, the pairs separate in space. Asymptotically for long times, entropy of entanglement between any two half-planes grows linearly in time in proportion to the number of pairs separated by the border-line between the half-planes~\cite{quasiparticle_horizon}. Accordingly, the bond dimension would have to grow exponentially in order to accommodate all this entanglement. Therefore, ultimately the tensor network is in general expected to fail after a finite time with only logarithmic progress being possible by increasing $D$, even when the best possible use of the bond dimension is made. Our goal here is, therefore, not to overcome the quasiparticle horizon effect~\cite{quasiparticle_horizon} but to get closer to the optimal use of the bond dimension. 

Our present results are shown in Figs. \ref{fig:x}. As a benchmark we also show full update (FU) results \cite{CzarnikDziarmagaCorboz} with $D=8$ up to times where they appear converged with this bond dimension. We also include NTU results \cite{ntu} with $D=8$ up to time when their energy per site departs by $0.01$ from its initial value at $t=0^+$. Truncation errors quantified by the overlap \eqref{O} are shown in Figs. \ref{fig:O_NTU} and \ref{fig:O_SVDU} in appendix \ref{app:errors}. The GTU simulations are terminated when the energy departs by $0.01$ or the truncation error exceeds a threshold, whichever comes first.   

The quench to the quantum critical point, $h_x=h_c$, is more challenging. Much the same as for FU and NTU, progress in evolution time made by increasing $D$ is slow. Nevertheless, GTU evolution time achieved with $D=6$ is somewhat longer than for FU/NTU with $D=8$. Furthermore, the quench to $h_x=2h_c$ yields an even more promising result: GTU with $D=6$ increases the evolution time more than twice as compared to FU with $D=8$. This is not quite unexpected as less excitation is created at $h_x=2h_c$ and the excitation spectrum is gapfull making the quasiparticle pairs separate more slowly. 

\begin{figure}[t!]
\vspace{-0cm}
\includegraphics[width=1.0\columnwidth,clip=true]{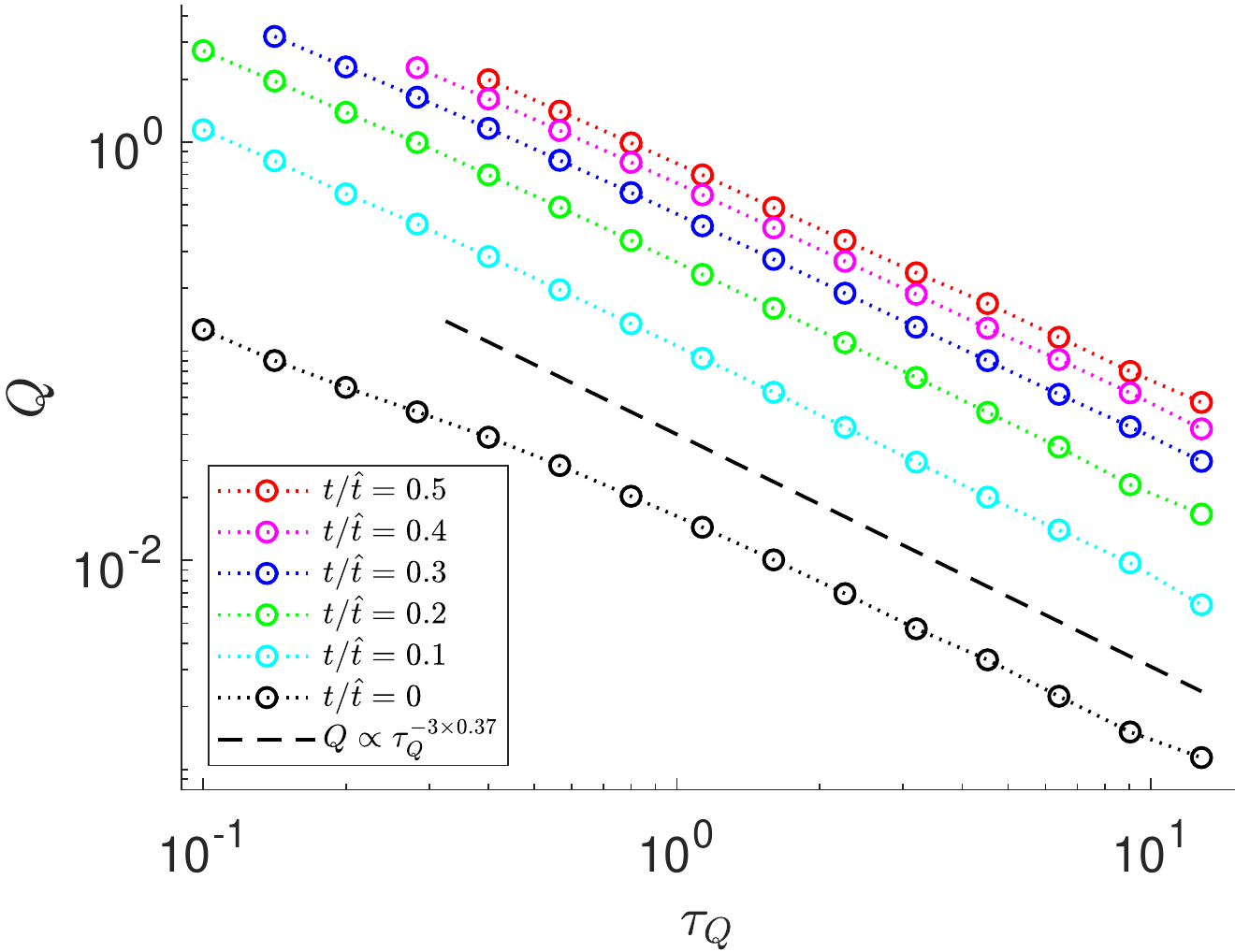}
\vspace{-0cm}
\caption{
{\bf Kibble-Zurek ramp.} 
The excitation energy per site, $Q$, in function of the quench time $\tau_Q$ during the ramp \eqref{eq:ramp_linear}. $Q$ is calculated at $t=0$, when the ramp is crossing the critical point, and at several later scaled times, $t/\hat t$. The scaling hypothesis \eqref{QKZ} implies a scaling $Q\propto\hat\xi^{-3}\propto\tau_Q^{-3\times 0.386}$ for each $t/\hat t$. As shown by the dashed line, $0.37$ in place of the exact $0.386$ is a better fit to our data for the longest achievable $\tau_Q$. 
Here all data were obtained with $D=3$.
}
\label{fig:kz}
\end{figure}

\section{Quantum Kibble-Zurek mechanism}
\label{sec:kz}

In this section instead of the sudden quench we consider a continuous ramp
\begin{eqnarray}
h_x(t)/h_c = 1-\epsilon(t),~~~ J(t) = 1+\epsilon(t). 
\label{eq:ramp_linear} 
\end{eqnarray}  
Here 
\be 
\epsilon(t) = 
\left\{
\begin{array}{ll}
\frac{t}{\tau_Q} - \frac{4}{27} \frac{t^3}{\tau_Q^3}, & {\rm when~~} t\leq 0 \\
\frac{t}{\tau_Q}, & {\rm when~~} t>0
\end{array}
\right.
\ee 
with the time running from $t_i=-\frac{3}{2}\tau_Q$ to $t_f = \tau_Q$. Near the critical point, at $t=0$, the ramp is linear $\epsilon(t)\approx t/\tau_Q$ with a quench time $\tau_Q$. Near the initial time it is bending in order to avoid a discontinuous time derivative that would generate some excitations already at the very beginning of the ramp that would have to be accounted for by extra bond dimension but which are not interesting from the point of view of the quantum KZ mechanism (QKZM)~\cite{K-a, K-b, K-c, Z-a, Z-b, Z-c, Z-d}. The latter quantifies excitations generated in the universal regime near the critical point where the evolution is bound to become non-adiabatic due to closing of the energy gap at the criticality~\cite{QKZ1,QKZ2,QKZ3,d2005,d2010-a,d2010-b}. The KZ ramp in the 2D quantum Ising model was numerically simulated in Ref. \onlinecite{KZ2D}. First attempt at its quantum simulation with Rydberg atoms was made in Ref. \onlinecite{rydberg2d1}. 

One of the predictions of the QKZM is a scaling hypothesis for excitation energy per site, $Q$, as a function of time~\cite{KZscaling1,KZscaling2,Francuzetal}:
\be 
Q(t)=\hat\xi^{-(z+d)} F_Q\left( t/\hat t \right). 
\label{QKZ}
\ee 
Here $d=2$ is dimensionality of our 2D system, $z=1$ is its dynamical exponent, 
$\hat \xi\propto \tau_Q^{\nu/(1+z\nu)}$ is the KZ correlation length, where $\nu=0.629971$ is the correlation length exponent, $\hat t\propto\hat\xi^z$ is the KZ scale of time, and $F_Q$ is a non-universal scaling function. Eq. \ref{QKZ} is expected to be valid near the critical point for times between $-\hat t$ and $+\hat t$. In particular it implies $Q(t)\propto\hat\xi^{-(z+d)}=\hat\xi^{-3}\propto\tau_Q^{-3\nu/(1+z\nu)}=\tau_Q^{-3\times 0.386}$ for any fixed $t/\hat t$ in this regime.

With GTU we are testing this prediction in Fig. \ref{fig:kz} where $Q$ is plotted in function of $\tau_Q$ for several values of scaled time $t/\hat t =0,...,0.5$. Here we set $\hat t=1\tau_Q^{\nu/(1+z\nu)}=\tau_Q^{0.386}$ with a unit numerical prefactor. The actual $\hat t$ was estimated~\cite{KZ2D} as approximately one fourth of this value implying that the scaling hypothesis is expected to hold up to $t/\hat t\approx 0.25$ instead of $\approx 1$. The log-log plots in Fig. \ref{fig:kz} demonstrate that for the longest available $\tau_Q$ the data approach a power law $Q(t/\hat t)\propto \tau_Q^{-3\times 0.37}$. Although the quench times obtained here are $3..4$ times longer than in Ref. \onlinecite{KZ2D}, where the plain NTU was used for the same simulations, there are still appreciable non-universal corrections to the exact exponent making it closer to $0.37$ in place of $0.386$. For the longest $\tau_Q=12.8$ small extra oscillations on top of the KZ excitation energy are visible for smaller $t/\hat t$ where the KZ energy is still relatively small. They are induced by truncation of the bond dimension. This problem becomes more severe for even longer quench times. 

\section{Conclusion}
\label{sec:conclusion}

The gradient optimization can significantly increase evolution times achievable with iPEPS
as compared to more standard methods like the full update or the neighborhood tensor update,
both after a sudden quench and during a linear Kibble-Zurek ramp.
In this paper the corner transfer matrix renormalization group was employed to calculate the Gramm-Schmidt metric tensor and the gradient vector. Further progress may be achievable with the help of variational methods~\cite{vanderstraeten2021variational}.
The same gradient method could also be applied in imaginary time evolution simulating thermal states.

\acknowledgements
%
This work was 
funded by National Science Centre (NCN), Poland under project 2019/35/B/ST3/01028. 
%

\bibliography{ref.bib} 
\appendix

\begin{figure}[t!]
\vspace{-0cm}
\includegraphics[width=0.9\columnwidth,clip=true]{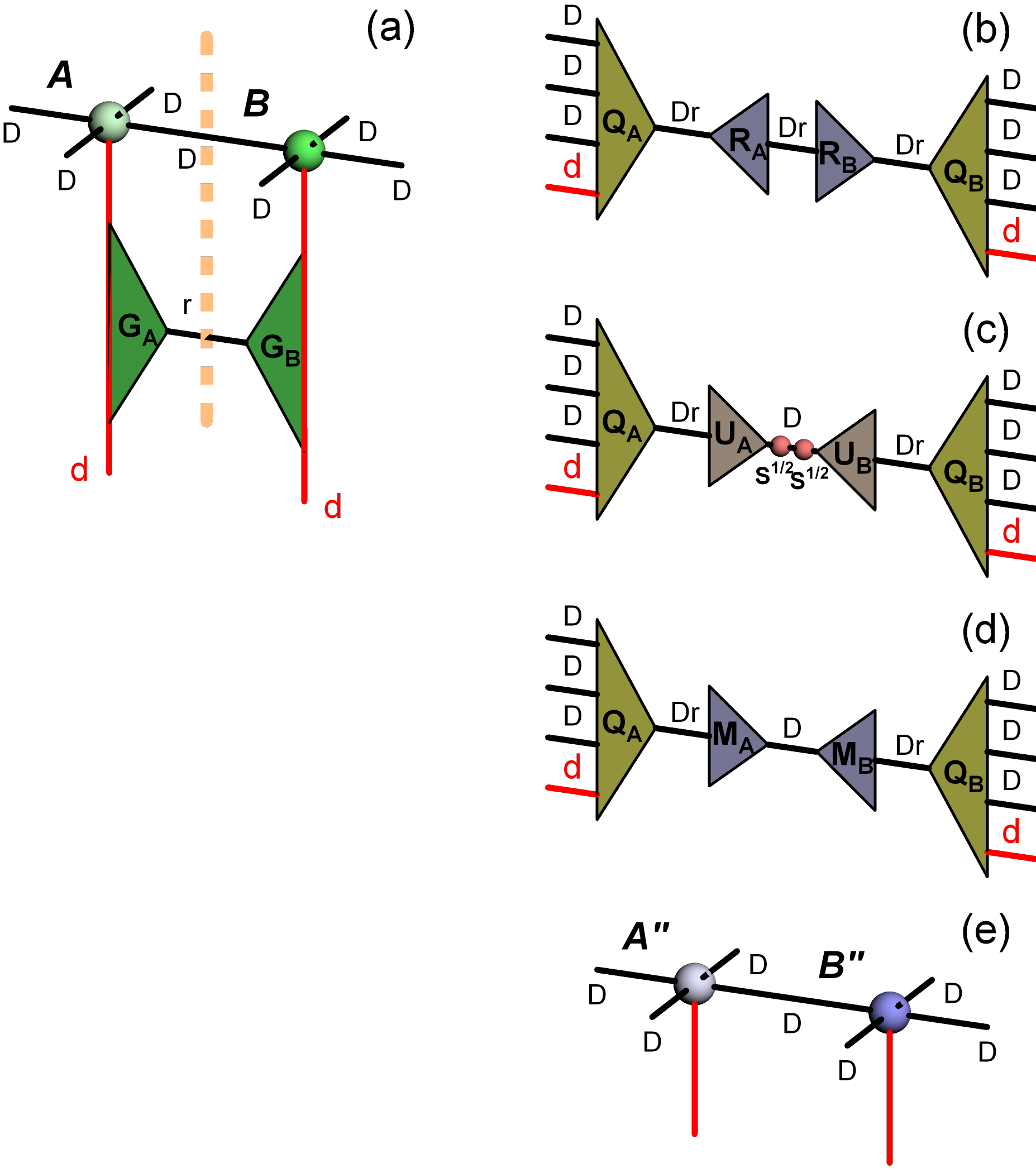}
\vspace{-0cm}
\caption{
{\bf Reduced tensors. }
In (a) 
a 2-site gate is applied to physical indices of NN tensors $A$ and $B$ as in Fig. \ref{fig:2site} (b). The gate is made of two tensors, $G_A$ and $G_B$, contracted by an index of dimension $r$.
In (b)
the tensor contraction $A\cdot G_A$ is QR-decomposed into $Q_AR_A$. Similarly $B\cdot G_B=Q_BR_B$. 
Isometries $Q_{A,B}$ will remain fixed until the Trotter gate is completed.
In (c)
after SVD, $R_AR_B^T=U_ASU_B^T$, $S$ is truncated to $D$ leading singular values.
In (d)
matrices $M_A=U_AS^{1/2}$ and $M_B^T=S^{1/2}U_B^T$ are made by absorbing a square root of truncated $S$ symmetrically.
They are the reduced tensors to be optimized.
In (e)
at this point one could abstain from further optimization and make new iPEPS tensors as $A^{''}=Q_A\cdot M_A$ and $B^{''}=Q_B\cdot M_B$ completing the Trotter gate. This scheme was referred to as SVD update (SVDU) in Ref. \onlinecite{ntu}. 
In NTU scheme reduced matrices $M_{A,B}$ are further optimized in the neighborhood tensor environment in Fig. \ref{fig:ntu}. 
In GTU the NTU optimized $M_{A,B}$ are further optimized by the gradient method before being contracted back with the fixed isometries $Q_{A,B}$ to make new $A^{''}$ and $B^{''}$.
}
\label{fig:reduced}
\end{figure}
\section{Reduced tensors}
\label{app:reduced}

The gradient optimization in Sec. \ref{sec:gradient} is not performed on full tensors $A^{''}$ and $B^{''}$ but on their much smaller reduced tensors/matrices $M_A$ and $M_B$ defined in Fig. \ref{fig:reduced}. They are the actual variational parameters optimized by the gradient method. The reduction makes the metric tensor and the gradient much more compact. For instance, the metric is not a $D^4 d\times D^4 d$ matrix but instead a $D^2 d \times D^2 d$ one. Not only the final metric is smaller but, more importantly, tensors involved in its computation by CTM, which is the bottleneck of the method, are more compact by a factor of $D^2$.

\begin{figure}[t!]
\vspace{-0cm}
\includegraphics[width=1.0\columnwidth,clip=true]{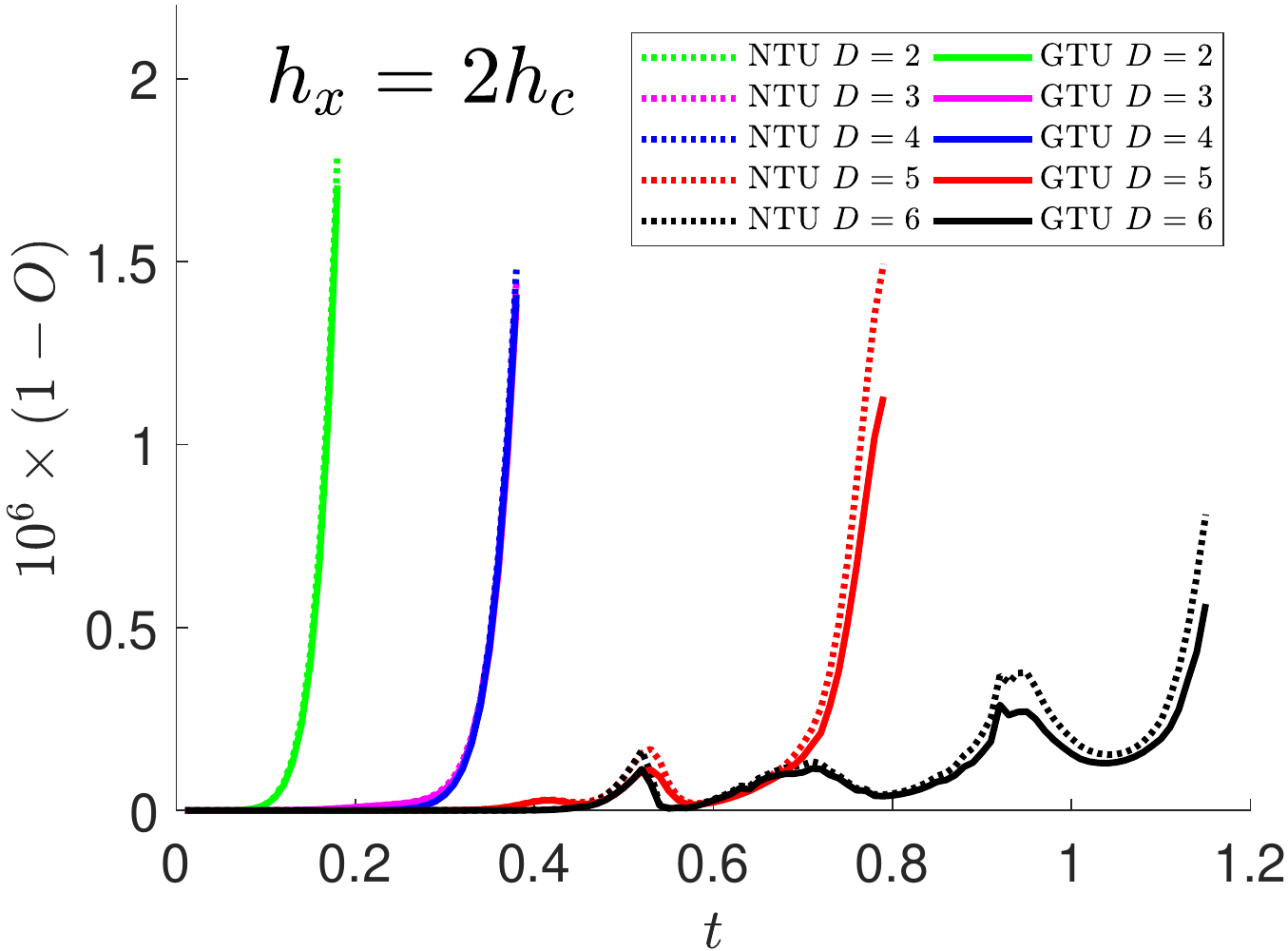}
\includegraphics[width=1.0\columnwidth,clip=true]{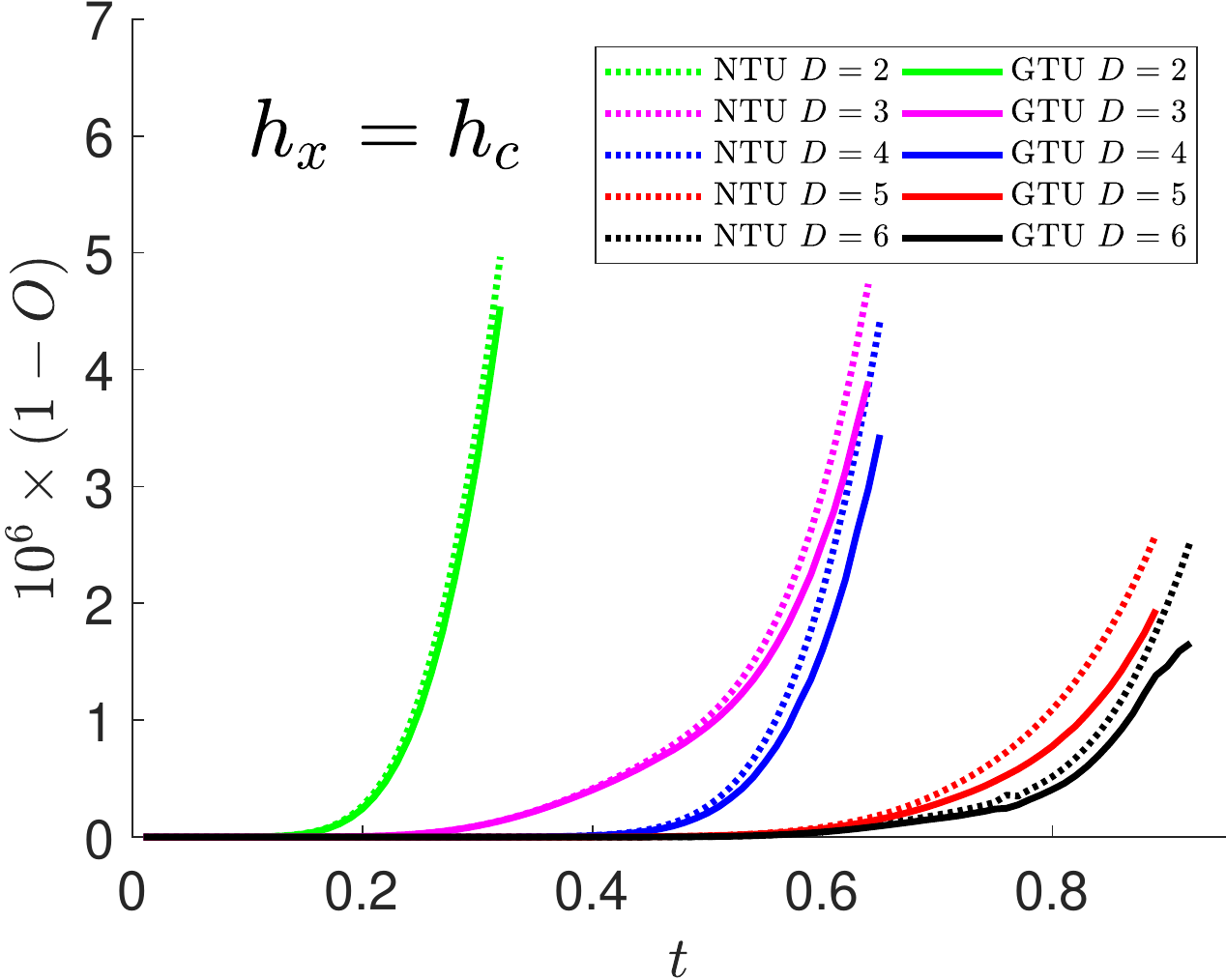}
\vspace{-0cm}
\caption{
{\bf Truncation error: GTU v. NTU.} 
The error is measured by $1-O$, where $O$ is the overlap per site between the exact and the truncated iPEPS in \eqref{O}. Here we show the error after NTU initialization (dotted) and the one after convergence of GTU optimization (solid). In case of $h=2h_c$ (top) the stopping criterion is GTU error greater than $2\times 10^{-6}$. In case of $h=h_c$ (bottom) it is $5\times10^{-6}$. 
}
\label{fig:O_NTU}
\end{figure}

\begin{figure}[t!]
\vspace{-0cm}
\includegraphics[width=1.0\columnwidth,clip=true]{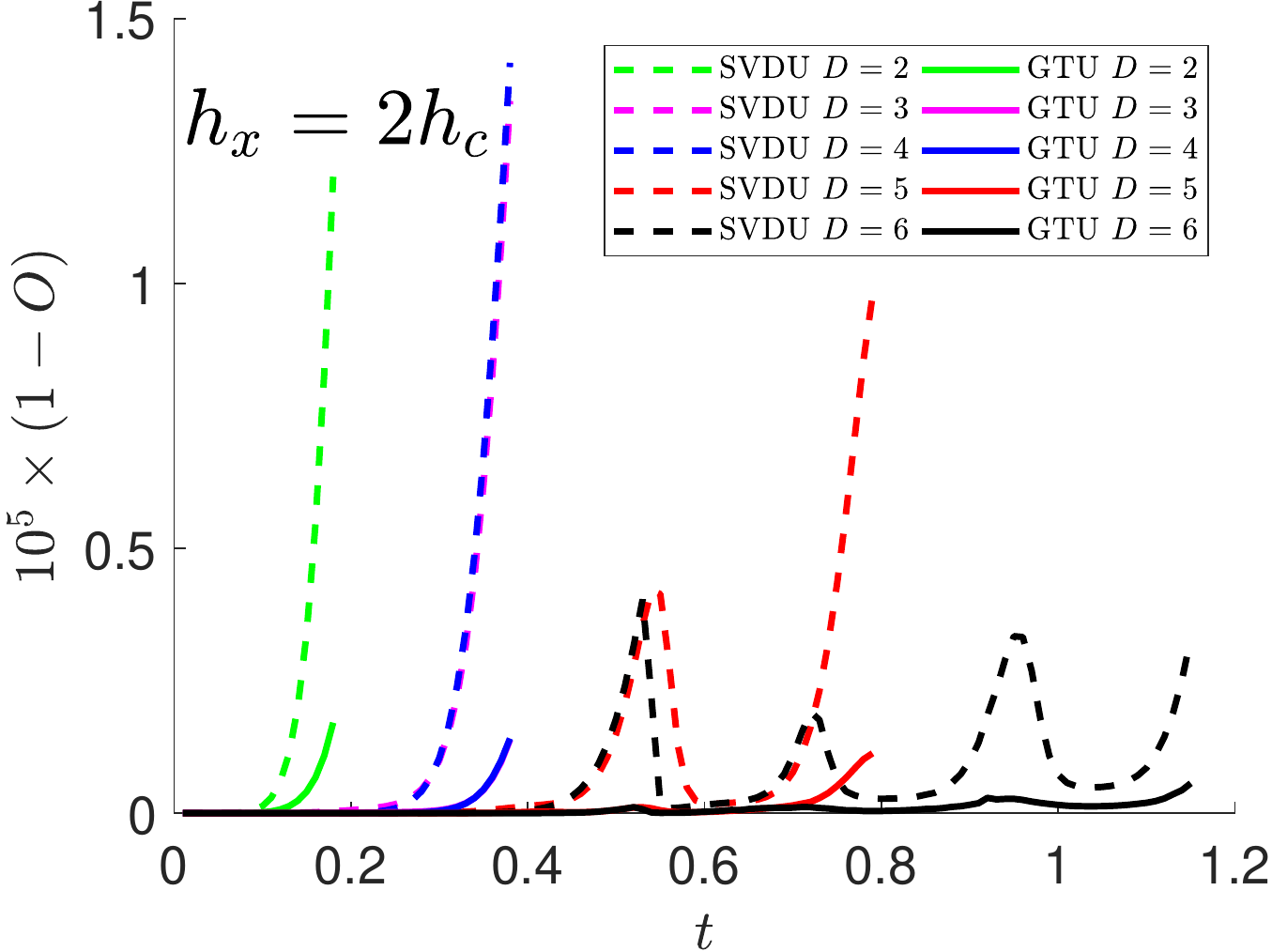}
\includegraphics[width=1.0\columnwidth,clip=true]{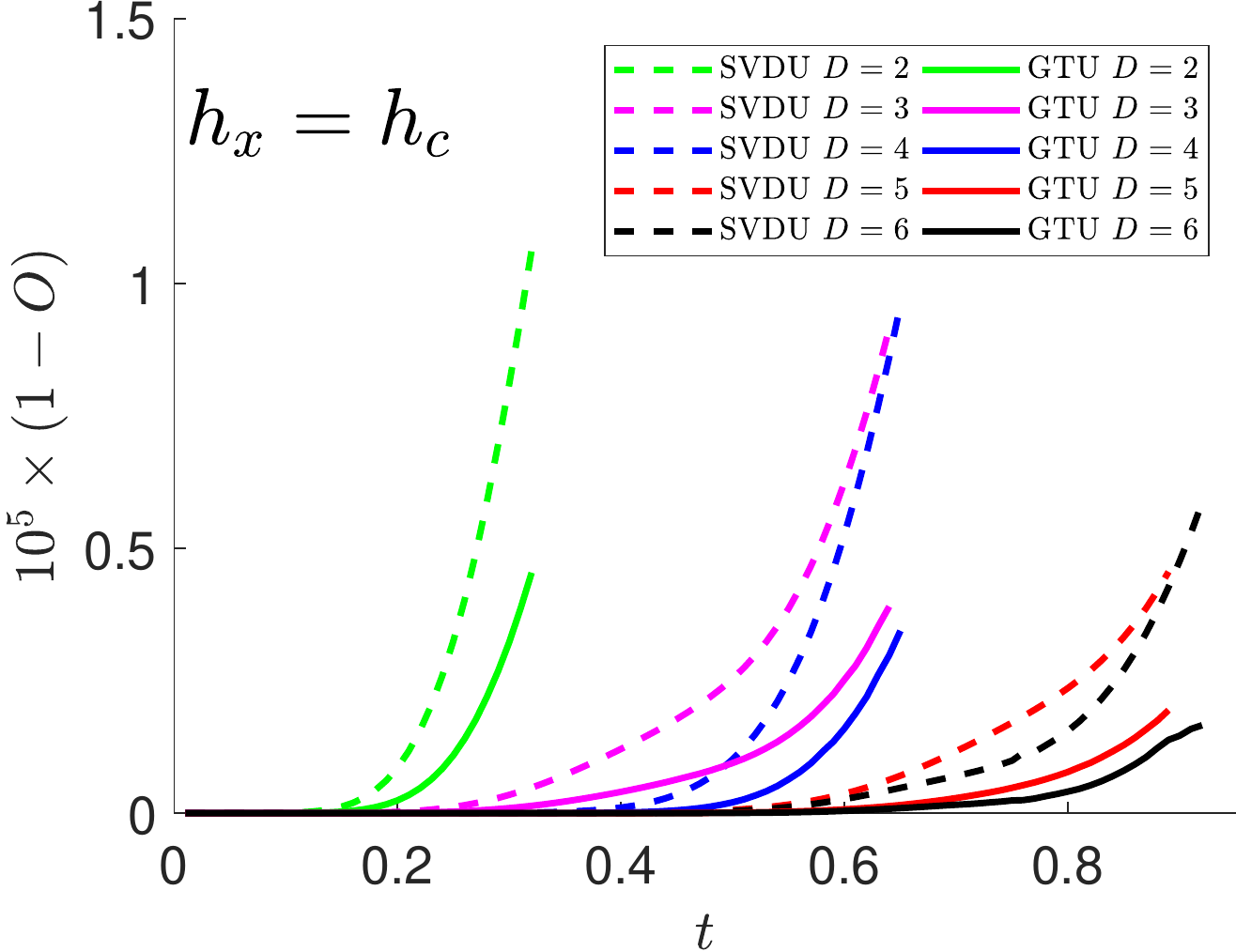}
\vspace{-0cm}
\caption{
{\bf Truncation error: GTU v. SVDU.} 
The error is measured by $1-O$, where $O$ is the overlap per site between the exact and the truncated iPEPS in \eqref{O}. Here we show the error after the SVDU initialization (dashed) and the one after convergence of GTU optimization (solid). 
}
\label{fig:O_SVDU}
\end{figure}

\section{Truncation errors}
\label{app:errors}

Figures \ref{fig:O_NTU} and \ref{fig:O_SVDU} show the error defined in \eqref{O} after, respectively, the NTU optimization and the SVDU initialization. Combining data in the figures we can infer that the error drops several times during the NTU optimization stage. The following GTU optimization can add another $20-30\%$ reduction.

\end{document}